\documentclass[%
 reprint,
 amsmath,amssymb,
 aps,
 prl,
]{revtex4-2}

\usepackage{graphicx}
\usepackage{dcolumn}
\usepackage{bm}
\usepackage{bbm}

\usepackage[utf8]{inputenc}  
\usepackage[T1]{fontenc}     
\usepackage{tensor}
\usepackage{amsmath}
\usepackage{amssymb}
\usepackage{graphicx}
\usepackage{mathrsfs}
\usepackage{amsfonts}
\usepackage{amsthm}
\usepackage{color}
\usepackage{txfonts}
\usepackage[colorlinks=true,citecolor=blue,linkcolor=blue,urlcolor=blue,anchorcolor=blue]{hyperref}%
\usepackage{pifont}
\hypersetup{colorlinks=true,citecolor=blue,linkcolor=blue,urlcolor=blue}
\usepackage[colorlinks=true,citecolor=blue,linkcolor=blue,urlcolor=blue,anchorcolor=blue]{hyperref}%
\providecommand{\U}[1]{\protect\rule{.1in}{.1in}}
\makeatletter
\@ifundefined{textcolor}{}
{
\definecolor{BLACK}{gray}{0}
\definecolor{WHITE}{gray}{1}
\definecolor{RED}{rgb}{1,0,0}
\definecolor{GREEN}{rgb}{0,1,0}
\definecolor{BLUE}{rgb}{0,0,1}
\definecolor{CYAN}{cmyk}{1,0,0,0}
\definecolor{MAGENTA}{cmyk}{0,1,0,0}
\definecolor{YELLOW}{cmyk}{0,0,1,0}
}

\begin{document}


\title{Current-driven nonlinear skyrmion dynamics in altermagnets}
\author{Yang Liu}
\author{Zhejunyu Jin}
\author{Jie Liu}
\author{Peng Yan}
\email[Contact author: ]{yan@uestc.edu.cn}  
\affiliation{School of Physics and State Key Laboratory of Electronic Thin Films and Integrated Devices, University of Electronic Science and Technology of China, Chengdu 610054, China}

\begin{abstract}
The center of mass and helicity are two dynamic degrees of freedom of skyrmions. In this work, we study the current-driven skyrmion motion in frustrated altermagnets. Contrary to conventional wisdom, we find that the skyrmion helicity is not locked with the skyrmion Hall angle, but unidirectionally rotates with a global angular velocity proportional to the square of the current density. In addition, we find that the helicity rotation velocity is highly anisotropic, depending on the direction of current flows. We also observe helicity oscillation in the terahertz regimes, where the nonlinear mixing between the fast and slow modes generates a magnon frequency comb. Full atomistic spin dynamics simulations verify our theoretical predictions. Our results establish frustrated altermagnets as a promising platform for skyrmionics, THz technology, and frequency comb.
\end{abstract}

\maketitle



\textit{Introduction}---Magnetic skyrmions are quasiparticles with nanoscale dimensions and topological stability \cite{Rossler2006, Nagaosa2013, 10.1126/science.aaa1442, Jiang2017, PhysRevLett.116.147203, Litzius2017, Legrand2020, RevModPhys.94.035005}, positioning them as promising carriers for next-generation information industry. Their swirling spin textures enable ultralow power for high-density storage \cite{Fert2013, Fert2017, Yu2017, Chen2024}, reconfigurable logic \cite{PhysRevB.94.054408, Luo2018, PhysRevApplied.12.064053, PhysRevB.107.054437, PhysRevApplied.22.054047}, and unconventional computing \cite{Song2020, PhysRevB.109.174420, lq2d-s6zm}. Beyond the center of mass, skyrmions host another helicity degree of freedom, the sense of spin rotation around the core \cite{Zhang2017, PhysRevB.97.224427, PhysRevB.111.064429}, that unlocks an exotic information channel. Translational motion of center-of-mass provides the ``orbital'' dynamics, while helicity rotation serves as the ``spin'' \cite{PhysRevB.93.064430}, paving the way for skyrmion qubits \cite{PhysRevLett.127.067201, PhysRevLett.130.106701, LiuPRB2024,PhysRevResearch.6.023067,PanPRL2024,SticletPRB2025}. Mastering helicity control is thus essential to propel skyrmion technologies from classical bits to quantum devices.
\setlength{\parskip}{0pt} 

\begin{figure}[htbp]
    \centering
    \includegraphics[width=0.48\textwidth]{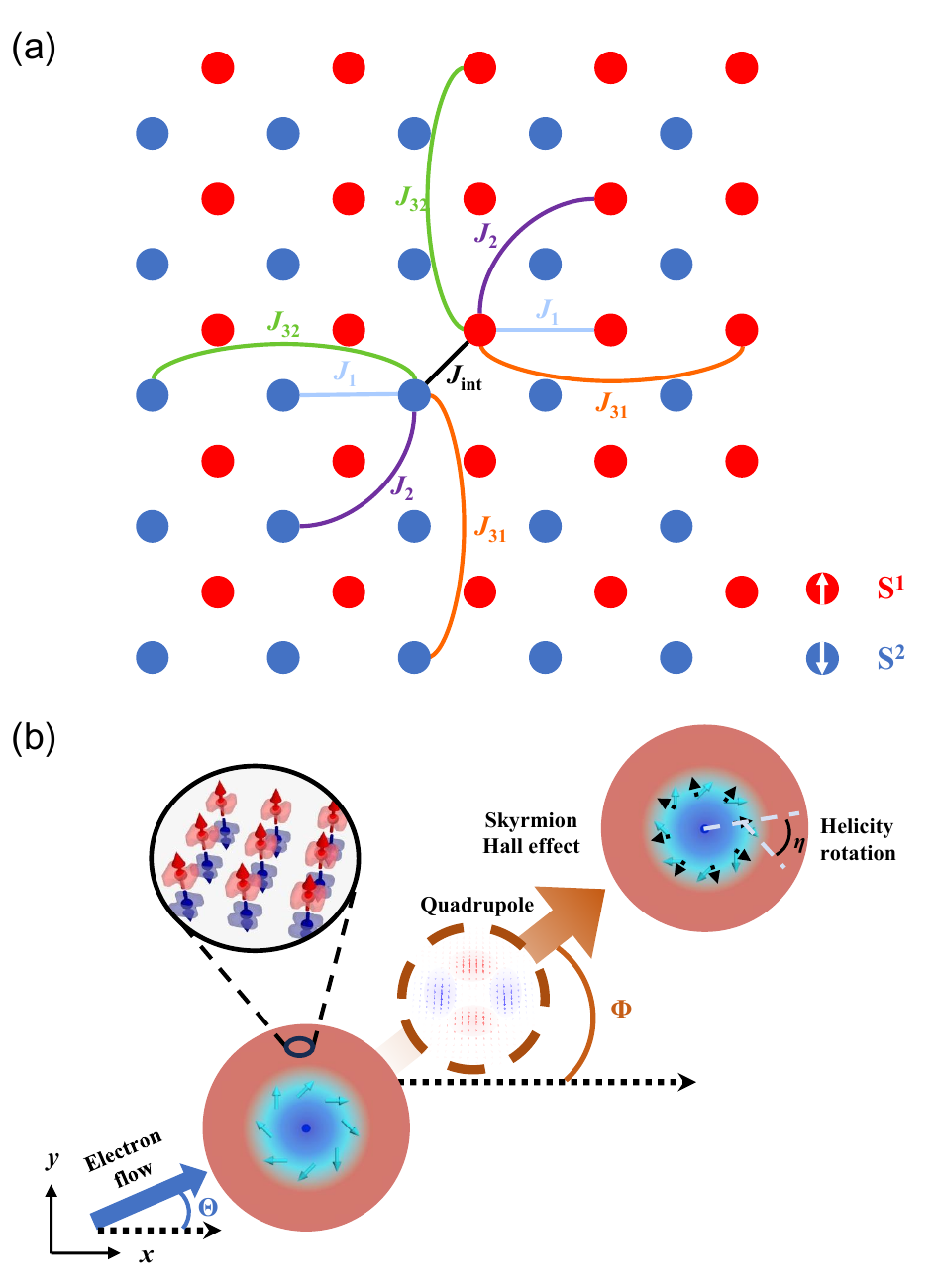}
    \caption{(a) Model of bilayer frustrated altermagnet with spin directions $\mathbf S^{n}$ ($n=1,2$). Anisotropic exchange interactions $J_{31}$ and $J_{32}$ act within each sublattice and obey twofold symmetry (green/orange bonds). One layer is shifted by $\sqrt{2}a/2$ along the diagonal line for clarity.
    (b) Schematic of altermagnetic skyrmion motion under electron flow. Adiabatic STT induces transverse Hall motion and nonlinear helicity ($\eta$) rotation assisted by the emergent magnetic quadrupole (dashed brown circle). $\Theta$ and $\Phi$ denote the angles of current and skyrmion velocity with respect to the $\hat{x}$-direction; red and blue arrows indicate alternating sublattice magnetizations.}
    \label{fig1}
\end{figure}

However, the manipulation of skyrmion helicity remains highly challenging. In frustrated ferromagnets, the dipole-dipole interaction (DDI) sets an energy barrier that prevents the helicity rotation \cite{PhysRevB.106.054414}. In relativistic ferromagnets or antiferromagnets, the chiral Dzyaloshinskii-Moriya interaction (DMI) locks a fixed helicity \cite{PhysRevLett.127.117204, Niu2025, Srivastava2018,FullertonAdvM2025,Msiska2022}. Altermagnets, emerging magnets with nonrelativistic spin splitting \cite{PhysRevX.12.031042, PhysRevX.12.040501,Song2025, Jiang2025, sciadv.aaz8809, PhysRevB.102.014422, PhysRevB.102.144441, Zhu2024, PhysRevLett.131.256703, PhysRevLett.126.127701, Smejkal2022, PhysRevB.108.L180401, PhysRevLett.130.036702, PhysRevLett.126.127701,Bose2022, PhysRevLett.128.197202, PhysRevLett.129.137201}, are promising to overcome this obstacle. It has been shown that altermagnets can host noncollinear magnetization textures \cite{Amin2024}. Sublattice canting induces finite local magnetic moments \cite{Gomonay2024}, yielding stripe domains and/or skyrmions with emergent magnetic quadrupoles \cite{PhysRevLett.133.196701, PhysRevLett.134.176401}. This drives unique magnetization dynamics, including anisotropic Walker breakdown \cite{Gomonay2024} and skyrmion Hall effect \cite{PhysRevLett.133.196701, PhysRevLett.134.176401} in altermagnets. While DMI has often been used to stabilize skyrmions \cite{PhysRevLett.133.196701, PhysRevLett.134.176401,JiangPRB2025,Dou2025arXiv}, its relativistic nature contrasts with the nonrelativistic feature of altermagnets, motivating alternative stabilization mechanisms. Frustrated interactions enable DMI-free skyrmion stabilization via competing exchanges and easy-axis anisotropy \cite{PhysRevLett.117.157205, PhysRevLett.108.017206, PhysRevLett.134.186701}, therefore offering a scalable route toward skyrmionics in the landscape of altermagnetism.

In this Letter, we study the current-driven skyrmion dynamics in frustrated altermagnets. By deriving the equations of motion for the coupled dynamics of skyrmion's center-of-mass and helicity, we show that the non-adiabatic spin-transfer-torque (STT) governs the linear longitudinal skyrmion propagation, while the adiabatic STT, coupled to the magnetic quadrupole, induces both the transverse drift and global helicity rotation with angular velocity proportional to the square of the current density. The helicity rotation is found to be anisotropic, depending on the current direction. The nonlinear mixing between the fast and slow helicity rotations generates the terahertz (THz) magnon frequency comb. Full atomistic spin dynamics simulations verify our theoretical predictions. These findings position altermagnets as an ideal platform for nonlinear helicity control for advanced skyrmionics.

\textit{Model}---We consider a bilayer model of altermagnets on a square lattice with the following Hamiltonian
\begin{subequations}\label{eq1}
\begin{align}
\mathcal{H} & = J_1 \sum_{\langle i, j \rangle} \mathbf{S}_i^n \cdot \mathbf{S}_j^n + J_2 \sum_{\langle \langle i, j \rangle \rangle} \mathbf{S}_i^n \cdot \mathbf{S}_j^n + J_{\text{int}} \sum_i \mathbf{S}_i^1 \cdot \mathbf{S}_i^2 \notag \\
& \quad - K \sum_i \left( \mathbf{S}_i^{n,z} \right)^2 + J_{31} \sum_{\langle \langle \langle i_x, j_x \rangle \rangle \rangle} \mathbf{S}_i^1 \cdot \mathbf{S}_j^1 + J_{32} \sum_{\langle \langle \langle i_y, j_y \rangle \rangle \rangle} \mathbf{S}_i^1 \cdot \mathbf{S}_j^1 \notag \\
& \quad + J_{32} \sum_{\langle \langle \langle i_x, j_x \rangle \rangle \rangle} \mathbf{S}_i^2 \cdot \mathbf{S}_j^2 + J_{31} \sum_{\langle \langle \langle i_y, j_y \rangle \rangle \rangle} \mathbf{S}_i^2 \cdot \mathbf{S}_j^2,
\tag{1}
\end{align}
\end{subequations}
where $J_1<0$ is the nearest-neighbor ferromagnetic exchange coupling, $J_2>0$ is the next-nearest-neighbor antiferromagnetic exchange interaction, $J_\mathrm{int}$ > 0 is the interlayer antiferromagnetic exchange constant, $J_{31}$ and $J_{32}$ are the anisotropic next-next-nearest-neighbor exchange interactions, and $K$ is the perpendicular magnetic anisotropy constant, as shown in Fig. \ref{fig1}(a). $\langle \langle \langle i_{x}, j_{x}\rangle \rangle \rangle$ and $\langle \langle \langle i_{y}, j_{y}\rangle \rangle \rangle$ indicate the next-next-nearest neighbor along the $\hat{x}$- and $\hat{y}$-direction, respectively. $\mathbf{S}_i^n$ represents the normalized spin at the site $i$ with $n=1,2$ being the layer index. By introducing the net spin $\mathbf{m}=(\mathbf{S}^1+\mathbf{S}^2)/2$ and Néel vector $\mathbf{n}=(\mathbf{S}^1-\mathbf{S}^2)/2$ \cite{PhysRevB.93.104408}, and expanding the Hamiltonian to the quartic order \cite{PhysRevB.93.064430, PhysRevLett.116.187202}, we obtain the continuum Ginzburg–Landau energy (see Supplemental Material \cite{SM} for detailed derivations)
\begin{subequations}\label{eq2}
\begin{align}
\mathcal{H} & = \int d\mathbf{r}\biggl\{I_1(\nabla\mathbf{n})^2 - J_2 \partial_x\mathbf{n} \cdot \partial_y\mathbf{n} + \frac{2J_2}{3}a^2(\partial_x\mathbf{n} \cdot \partial_y\mathbf{n})^2 \notag \\
& \quad + I_2a^2(\nabla^2\mathbf{n})^2  - I_3 (\partial_x\mathbf{m} \cdot \partial_x\mathbf{n} - \partial_y\mathbf{m} \cdot \partial_y\mathbf{n}) + 2J_{\text{int}}\mathbf{m}^2 \notag \\
& \quad + \frac{I_3}{3}a^2 (\partial_x^2\mathbf{m} \cdot \partial_x^2\mathbf{n}-\partial_y^2\mathbf{m} \cdot \partial_y^2\mathbf{n}) - 2K n_z^2\biggr\},
\tag{2}
\end{align}
\end{subequations}
with $I_1 = - (J_1 + J_2 + 2J_{31} + 2J_{32})/2$, $I_2 =(J_1 + 2J_2 + 8J_{31} + 8J_{32})/{12}$, and $I_3 = 4(J_{31} - J_{32})$. Here, $a$ represents the lattice constant. The Lagrangian is given by \cite{RevModPhys.90.015005,PhysRevLett.131.166704,PhysRevLett.134.116701}
\begin{equation}\label{eq3}
\mathcal{L} = 2s\int{d\mathbf{r} \biggl\{\left(\partial_{t}\mathbf{n} \times \mathbf{n}\right) \cdot \mathbf{m}} \biggl\}~-~\mathcal{H},
\end{equation}
where $s = \mu_s/\gamma$ is the spin angular momentum with the atomic magnetic moment $\mu_s$ and gyromagnetic ratio $\gamma$. Utilizing the Euler–Lagrange equation, we have
\begin{equation}\label{eq4}
\mathbf{m} = \frac{s}{2J_{\text{int}}} (\partial_{t}\mathbf{n} \times \mathbf{n} ) - \mathbf{A},
\end{equation} 
with the vector potential
\begin{subequations}\label{eq5}
\begin{align}
\mathbf{A} & = \frac{I_3}{2J_{\text{int}}} a^{2}\mathbf{n} \times \big [( \partial_x^2\mathbf{n}-\partial_y^2 \mathbf{n} ) \times \mathbf{n} \big ]. \tag{5}
\end{align}
\end{subequations}
Here, we have calculated the variational derivatives $\delta_ \mathbf{m}\mathcal{L}$ by varying $\mathbf{m}$ normal to a fixed $\mathbf{n}$, to enforce the constraints $|\mathbf{n}| = 1$ and $\mathbf{m} \cdot \mathbf{n} = 0$ \cite{PhysRevLett.106.107206, nlgc-rh7s}. It is noted that an isolated skyrmion forms spontaneously by choosing suitable materials parameters (see Fig. S1 \cite{SM}). Figure \ref{fig1}(b) shows the skyrmion stabilized under frustrated exchange parameters $J_{1}=-6$ meV, $J_{2}=1.2$ meV, $J_{31}=0.9$ meV, $J_{32}=0.3$ meV, and anisotropy $K=0.12$ meV. We set the lattice constant $a=4.12$ $\AA$ in atomistic spin dynamics simulations. The finite $\mathbf{m}$ forms a magnetic quadrupole (see Fig. \ref{fig1} and Sec. III \cite{SM}), assisting the nonlinear skyrmion dynamics driven by STT, as analyzed below.


\textit{STT-induced altermagnetic skyrmion dynamics}---In antiferromagnets, compensated skyrmions produce vanishing Magnus force or helicity rotation \cite{PhysRevLett.116.147203, science.add5751}. In altermagnets, sublattice canting generates finite local magnetic moments and emergent gauge field [Eq. (\ref{eq5})]. By including the STT, the coupled equations of motion for both magnetization $\mathbf{m}$ and Néel vector $\mathbf{n}$ read
\begin{subequations}\label{eq6}
\begin{equation}\label{eq6a}
  \dot{\mathbf{m}} = - \frac{1}{2s}\mathbf{n} \times \mathbf{f}_{\mathbf{n}} + \alpha\mathbf{n}\times \dot{\mathbf{n}} + \left( \mathbf{u} \cdot \nabla\right)\mathbf{m} - \beta \mathbf{n} \times \left( \mathbf{u} \cdot \nabla\right)\mathbf{n},
\end{equation} 
\vspace{-2em} 
\begin{equation}\label{eq6b}
  \dot{\mathbf{n}} = - \frac{1}{2s} \mathbf{n} \times \mathbf{f}_{\mathbf{m}} + \left( \mathbf{u} \cdot \nabla\right)\mathbf{n},
\end{equation}
\end{subequations}
where $\mathbf{f_n} = \delta \mathcal{H}/\delta \mathbf{n}$ and $\mathbf{f_m} = \delta \mathcal{H}/\delta \mathbf{m}$ denote the effective fields of $\mathbf{n}$ and $\mathbf{m}$, respectively, $\alpha$ is the Gilbert damping constant, and $\mathbf{u} = -Pj_ca^3/e(1 + \beta^2)\mathbf{j}$ is the drift velocity of conduction electrons, with the current density $j_c$, polarization $P$, the elementary charge $e$, non-adiabatic coefficient $\beta$, and the current direction $\mathbf{j}$ \cite{PhysRevLett.93.127204}. Assuming a rigid-body motion, we parameterize skyrmion by its guiding center $\mathbf{R}(t)$ and helicity $\eta(t)$: $\mathbf{n}(t) = \mathbf{n}_0[\mathbf{r} - \mathbf{R}(t), \eta(t)]$, where $\mathbf{n}_0(\mathbf{r})$ is the equilibrium skyrmion profile without current, and $\eta(t)$ sets the local spin's azimuthal angle $\varphi(t) = \phi + \eta(t)$, see Fig. \ref{fig1}(b), with $\phi$ denoting the spatial azimuthal angle \cite{Zhang2017,10.1063/5.0146374}. To model the effect from STT, we implement the transformation $\partial_t \to \partial_t+\mathbf{u} \cdot \nabla$ in the Lagrangian \cite{PhysRevB.94.054409}, yielding $\mathbf{n}=\mathbf{n}_0+\delta\mathbf{n}$ and $\mathbf{m}=\mathbf{m}_0+\delta\mathbf{m}$ with $\delta\mathbf{n}=(-s/2J_{\text{int}}) (\mathbf{u} \cdot \nabla)\mathbf{A} \times \mathbf{n} $ and $\delta\mathbf{m}=(s/2J_\text{int}) (\mathbf{u}\cdot\nabla) \mathbf{n} \times \mathbf{n}$. 

Since the velocity of skyrmion's center-of-mass $\mathbf{R}=(R_x, R_y)$ is much smaller than the group velocity of magnons in altermagnets, one can safely ignore the Lorentz contraction effect (see Sec.~IV \cite{SM}). We then obtain
\begin{subequations}\label{eq7}
\begin{align}
& \frac{\rho}{2s}\ddot{\mathbf{n}} \times \mathbf{n}  = \frac{1}{2s}\mathbf{n} \times \mathbf{f_n} - \alpha(\mathbf{n} \times \dot{\mathbf{n}}+\mathbf{A} \times \dot{\mathbf{A}}) + \frac{\rho}{s}\dot{\mathbf{n}} \times (\mathbf{u}\cdot \nabla)\mathbf{n} \notag \\
& \quad + (\partial_t - \mathbf{u}\cdot \nabla) \mathbf{A} - \frac{\rho}{2s}(\mathbf{u}\cdot \nabla)^{2} \mathbf{n} \times \mathbf{n}+\beta\mathbf{n} \times (\mathbf{u}\cdot \nabla) \mathbf{n},
\tag{7}
\end{align}
\end{subequations}
where $\rho = s^2/J_\mathrm{int}$ is the spin inertia. In terms of the collective-coordinate method \cite{PhysRevLett.30.230}, we derive the coupled equations of motion for altermagnetic skyrmions
\begin{subequations}\label{eq8}
\begin{equation}\label{eq8a}
\overset{\leftrightarrow}{\mathcal{M}}\ddot{\mathbf{R}} + \alpha\overset{\leftrightarrow}{\mathcal{G}}\dot{\mathbf{R}} + \alpha\dot{\eta}\mathcal{E}_{\eta} + \dot{\eta} \mathcal{F}_{\eta} - \overset{\leftrightarrow}{\mathcal{A}}\mathbf{u} - u\overset{\leftrightarrow}{\mathcal{B}}\mathbf{u} - \beta\overset{\leftrightarrow}{\mathcal{D}}\mathbf{u}=0,
\end{equation}
\vspace{-2em} 
\begin{equation}\label{eq8b}
\mathcal{I}\ddot{\eta} + \alpha\mathcal{G}_\eta \dot{\eta} - \frac{\partial V}{\partial \eta} - \mathcal{N}_\eta\cdot \mathbf{u} - \mathcal{A}\mathcal{S}_\eta  \cdot \mathbf{u} - \beta\mathcal{E}_\eta \cdot \mathbf{u}=0.
\end{equation}
\end{subequations}Here, $\mathcal{M}_{ij} = (\rho/2s)\int\partial_i\mathbf{n}\cdot\partial_j\mathbf{n}d\mathbf{r}$ is the mass tensor ($i,j=x,y$), $\dot{\mathbf{R}}(t)=(v_x,v_y)$ is the skyrmion's center-of-mass velocity, $\overset{\leftrightarrow}{\mathcal{G}}=\overset{\leftrightarrow}{\mathcal{D}}+\overset{\leftrightarrow}{\mathcal{Q}}$ is the dissipative tensor with $\mathcal{D}_{ij} = \int \partial_i \mathbf{n}\cdot\partial_j\mathbf{n}d\mathbf{r}$ and $\mathcal{Q}_{ij} = \int (\mathbf{A}\times\partial_i\mathbf{A})\cdot\partial_j\mathbf{n}d\mathbf{r}$. 
The third and forth terms in the left-hand side of Eq.~(\ref{eq8a}) arise from the coupling between the skyrmion drift and helicity rotation with $\mathcal{E}_{\eta,i} = \int\partial_i\mathbf{n} \cdot \partial_{\eta}\mathbf{n} d\mathbf{r}$ and $\mathcal{F}_{\eta,i} = (\rho/s)\int\mathbf{n}\times(\partial_{\eta}\mathbf{n} \times (\mathbf{u}\cdot \nabla)\mathbf{n}) \cdot \partial_{i}\mathbf{n}d\mathbf{r}$ ($i=x,y$). The fifth term represents the effective force from the magnetic quadrupole with the dimensionless tensor $\mathcal{A}_{ij}=\int(\mathbf{n}\times \partial_i\mathbf{A}) \cdot \partial_j\mathbf{n} d\mathbf{r}$ (see the tensor profile in Fig. S3 \cite{SM}), and the sixth term indicates the second-order gradient correction induced by the current-induced skyrmion modification with $\mathcal{B}_{ij} = (\rho/2s)\int\partial_i^2\mathbf{n} \cdot \partial_{j}\mathbf{n} d\mathbf{r}$ and $u=-Pj_ca^3/e(1 + \beta^2)$. The last term is the drag force from the nonadiabatic STT. In Eq.~(\ref{eq8b}), $\mathcal{I} = (\rho / 2s) \int (\partial_{\eta}\mathbf{n})^{2}\, d\mathbf{r}$ is the moment of inertia and $\mathcal{G}_\eta = \int \big [\partial_\eta \mathbf{n}\cdot\partial_\eta\mathbf{n} + (\mathbf{A} \times \partial_\eta\mathbf{A}) \cdot \partial_\eta\mathbf{n}\big ]d\mathbf{r}$ denotes the dissipative force associated with helicity rotation. $V(\eta)=I_3\int \partial_{\eta}(\partial_x\mathbf{A} \cdot \partial_x\mathbf{n} - \partial_y\mathbf{A} \cdot \partial_x\mathbf{n}) d\mathbf{r}$ is the potential originating from the anisotropic exchange. $\mathcal{N}_{\eta,i} = \int(\mathbf{n} \times \partial_i\mathbf{A}) \cdot \partial_{\eta}\mathbf{n} d\mathbf{r}$ and $\mathcal{S}_{\eta,i} = (\rho u_i/s\alpha\mathcal{G})\int \big[\mathbf{n} \times ( \partial_j \mathbf{n} \times \partial_i \mathbf{n})\big] \cdot \partial_{\eta}\mathbf{n} d\mathbf{r}$ ($i\neq j,~i=x,y$) with $\mathcal{G}=\mathcal{G}_{xx}=\mathcal{G}_{yy}$ describe the first- and second-order gradient force exerted by the net spin on the altermagnetic skyrmion helicity, respectively. $\mathcal{A}= \mathcal{A}_{xy}=\mathcal{A}_{yx}$ arises from the coupling between the magnetic quadrupole and STT. In addition, we have examined that $\mathcal{A}_{xx}=\mathcal{A}_{yy}=0$ (see Figs. S3 and S4 \cite{SM}). The adiabatic force in Eq.~(\ref{eq8a}) can therefore be simplified as $\overset{\leftrightarrow}{\mathcal A}\mathbf u
=\mathcal A\,u\,(\sin\Theta\,,\cos\Theta)$ (see Sec. VI \cite{SM}), which encodes the rotation symmetry imposed by the magnetic quadrupole. Due to the rotational symmetry of the skyrmion, the potential $V(\eta)$ is flat, i.e., independent of $\eta$. Tensor $\overset{\leftrightarrow}{\mathcal{B}}$ in Eq.~(\ref{eq8a}) and vector $\mathbf{\mathcal{N}}_{\eta,i}$ in Eq.~(\ref{eq8b}) are negligibly small due to the 180° rotational symmetry of $\mathbf{m}$, but they can cause higher-order effects (see below).

\begin{figure}[t]
    \centering
    \includegraphics[width=0.48\textwidth]{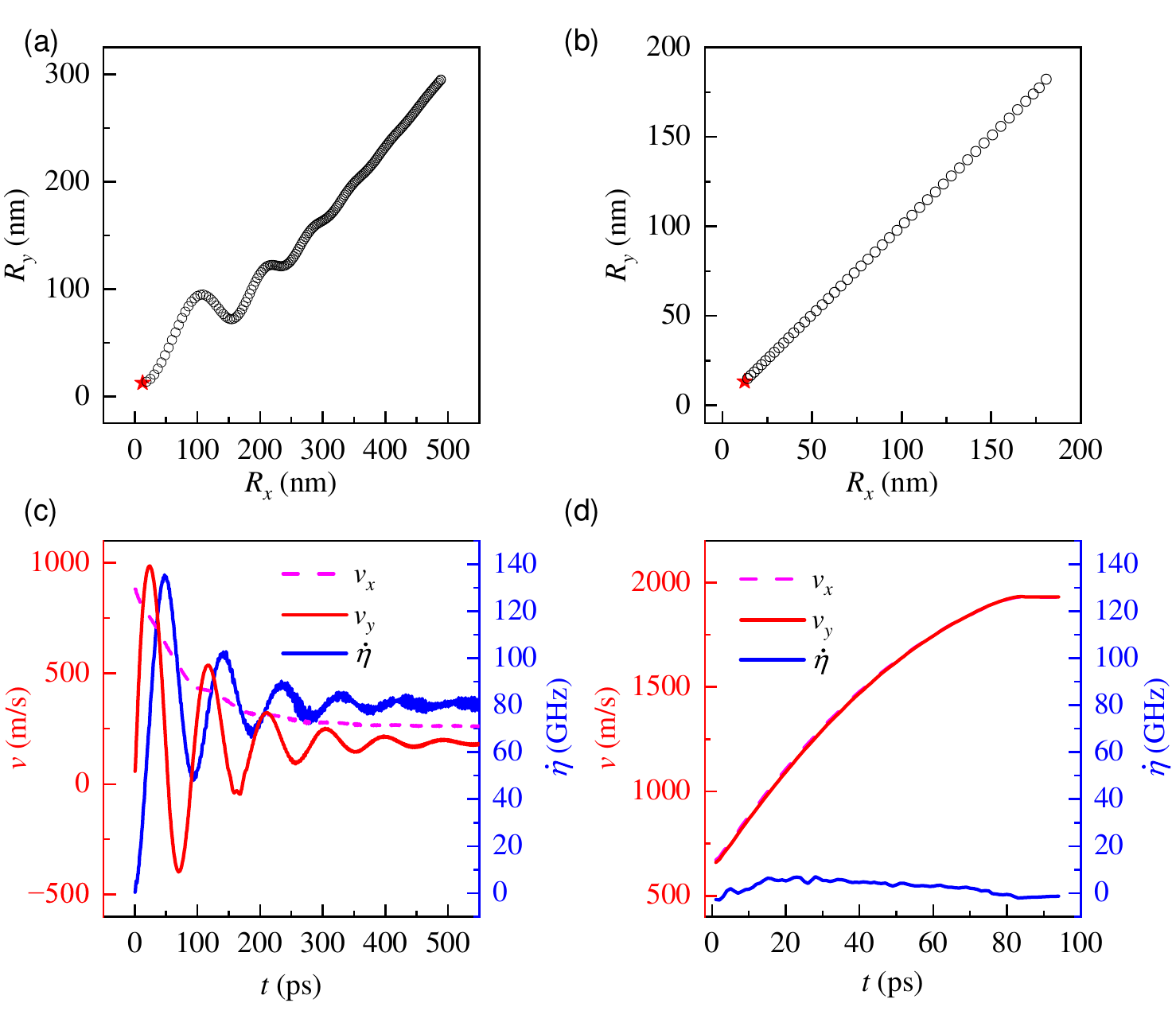}
    \caption{Trajectories of an altermagnetic skyrmion for current directions $\Theta = 0^\circ$ (a) and $45^\circ$ (b). The red stars indicates the initial position of the skyrmion center. Corresponding time-evolution of the skyrmion velocity and helicity rotation frequency for $\Theta = 0^\circ$ (c) and $45^\circ$ (d), respectively. Calculations are performed using parameters $u=1000~\mathrm{m/s}$, $\alpha=0.001$, and $\beta=0.0002$.}
    \label{fig2}
\end{figure}

Figure~\ref{fig2} plots the current-driven dynamics of an altermagnetic skyrmion. For a current injected along $\hat{x}$ ($\Theta = 0^\circ$), the skyrmion exhibits a pronounced wavy trajectory before relaxing into a steady motion, signaling a strong nonlinear effect, as shown in Fig.~\ref{fig2}(a). By contrast, when the current is applied at an oblique angle $\Theta = 45^\circ$, the skyrmion follows a straight trajectory [see Fig.~\ref{fig2}(b)]. The origin of this feature is further examined by tracing the time-evolution of both the skyrmion velocity and its helicity rotation. For $\Theta = 0^\circ$ [Fig.~\ref{fig2}(c)], both the transverse velocity $v_y$ and the helicity rotation frequency $\dot{\eta}$ strongly oscillate before reaching their steady-state values. Notably, the longitudinal velocity $v_x$ is decoupled from the helicity dynamics since $\mathcal{F}_{\eta,x}=0$, and it thus relaxes monotonically to a constant value without oscillations. In contrast, for $\Theta = 45^\circ$ [Fig.~\ref{fig2}(d)], the skyrmion's center-of-mass velocity increases smoothly and saturates finally, but the helicity velocity remains a constant. These results indicate that the nonlinear skyrmion dynamics originate from the coupling between skyrmion drift and helicity rotation, which was largely overlook in previous studies.

We then focus on the skyrmion's steady-state motion. Based on Eqs.~(\ref{eq8a}) and (\ref{eq8b}) and neglecting the inertia effects, we solve the skyrmion velocities
\begin{subequations}\label{eq9}
\begin{equation}\label{eq9a}
 \alpha\mathcal{G} v_x= \beta\mathcal{D}u_x + \mathcal{A}u_y,
\end{equation}
\vspace{-2em} 
\begin{equation}\label{eq9b}
 \alpha\mathcal{G} v_y= \mathcal{A}u_x + \beta\mathcal{D}u_y,
\end{equation}
\vspace{-2em} 
\begin{equation}\label{eq9c}
 \alpha\mathcal{G}_{\eta} \dot{\eta}= \frac{\rho\mathcal{A}\mathcal{S}}{s\alpha\mathcal{G} } (u_x^2 - u_y^2)+\beta(\mathcal{E}_{\eta,x} u_x+\mathcal{E}_{\eta,y} u_y),
\end{equation}
\end{subequations}which is the main result of this Letter. Here, $\mathcal{D}=\mathcal{D}_{xx}=\mathcal{D}_{yy}$, $\mathcal{S} = \int \big[\mathbf{n} \times ( \partial_y \mathbf{n} \times \partial_x \mathbf{n})\big] \cdot \partial_{\eta}\mathbf{n} d\mathbf{r}$, and $u_{x(y)}$ is the drift velocity of conduction electron along the $\hat{x}(\hat{y})$ direction. Equations (\ref{eq9}) therefore establish a Thiele framework for altermagnetic skyrmion motion under STT. Unlike isotropic DMI-stabilized skyrmions \cite{PhysRevLett.133.196701, PhysRevLett.134.176401}, frustrated altermagnets generate larger local magnetic moments, enhancing transverse motion via quadrupole tensor $\mathcal{A}_{ij}$, and thus unlocks the peculiar helicity rotation. Notably, the strong interlayer coupling $J_{\text{int}}$ enforces $\overline{\mathcal{E}_{\eta}(t)}=0$ in our model, which effectively decouples the dynamics of $\mathbf{R}$ from $\eta$ \cite{Leonov2015, Bernstein2025}. A fast oscillation of $\mathcal{E}_{\eta}(t)$ however persists (see below).

\begin{figure}[h]
    \centering
    \includegraphics[width=0.48\textwidth]{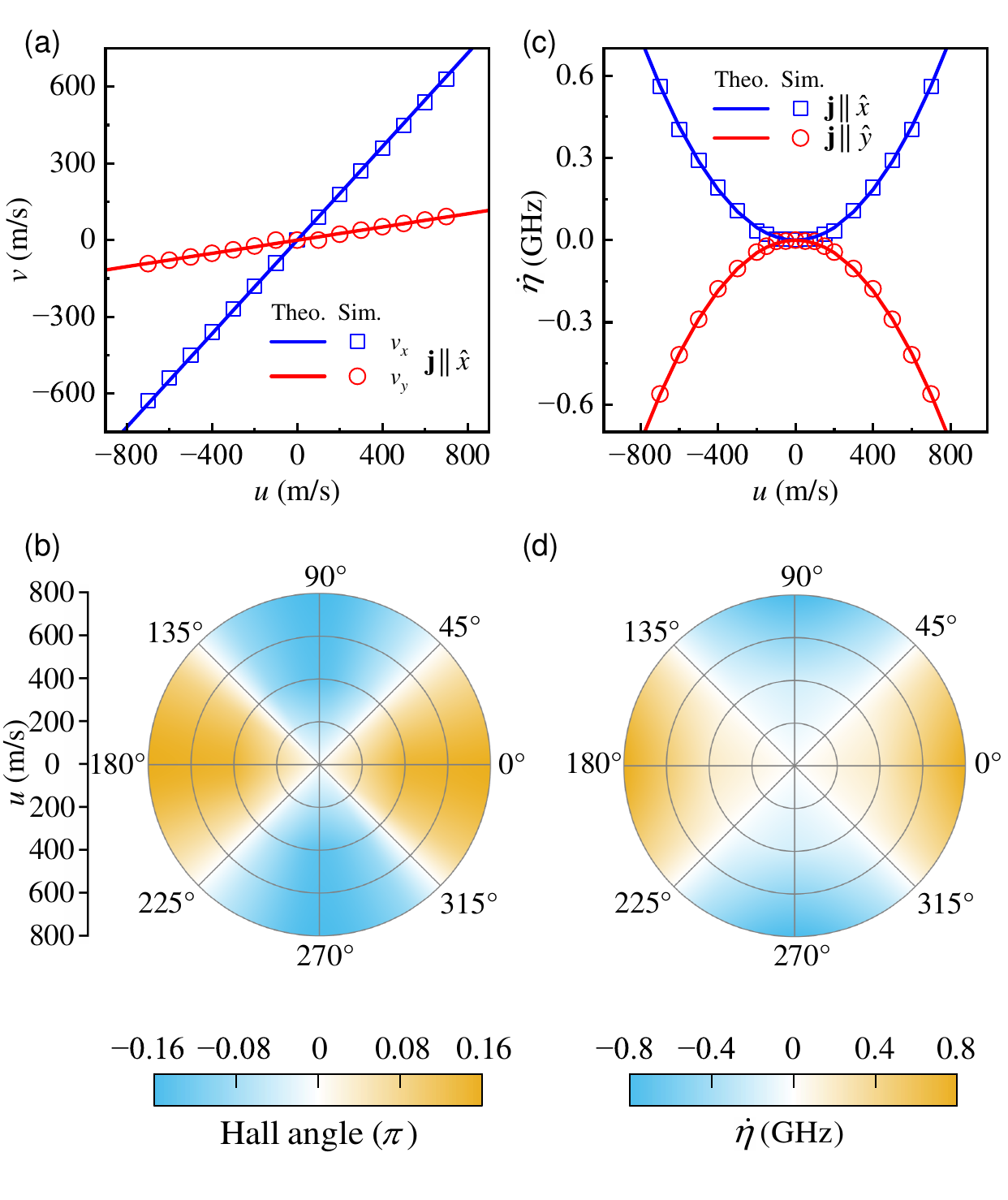}
    \caption{ Skyrmion propagation velocity (a) and helicity rotation frequency (c) as a function of the electron drift velocity $u$, with symbols from simulations and curves from Eqs. (\ref{eq9}). Skyrmion Hall angle (b) and helicity rotation frequency (d) vs. current direction (azimuthal: $\mathbf{j}$ angle) and electron drift velocity (radial: $u$ in units of m/s). In calculations, we set $I_{3}=2.4~\mathrm{meV}$, $\alpha=\beta=0.05$, and $J_{\mathrm{int}} = 4.2~\mathrm{meV}$.}
    \label{fig3}
\end{figure}

Figure \ref{fig3}(a) shows the skyrmion steady propagation velocity under current along $\hat{x}$. It is found that both $v_x$ and $v_y$ linearly increases with $u$, while $v_x$ (longitudinal) carries a steeper slope than $v_y$ (transverse), reflecting the quadrupole origin of the latter. Analytical formulas (curves) compare well with simulations (symbols), by adopting the numerically calculated parameters $\mathcal{G}=11.6$, $\mathcal{D}=10.6$, and $\mathcal{A}=0.075$, without any fitting. Figure~\ref{fig3}(b) shows the skyrmion Hall angle as a function of the current direction, where the Hall angle is defined as $(\Phi - \Theta)$ (see Fig.~\ref{fig1}). It is noted that the Hall effect vanishes at $45^{\circ}$, $135^{\circ}$, $225^{\circ}$, and $315^{\circ}$, where $\lvert v_x\rvert=\lvert v_y\rvert$, thus following a $d$-wave symmetry. Figure \ref{fig3}(c) reveals an interesting helicity dynamics: the global sub-gigahertz (GHz) helicity rotation $\dot{\eta}$ switches the sign from $\hat{x}$ to $\hat{y}$ but maintains the same sign for $\pm \hat{x}$ or $\pm \hat{y}$, with its magnitude proportional to the square of $u$. Our theory well explains these numerical findings with calculated parameters $\mathcal{G}_{\eta}=7.7 \times 10^{-18} ~\text{m}^2$ and $\mathcal{S}=0.86$ (see Fig. S4 \cite{SM}). Figure \ref{fig3}(d) maps the helicity rotation frequency vs. both the current direction and density, again evidencing the feature of the magnetic quadrupole.

\begin{figure}[t]
    \centering
    \includegraphics[width=0.48\textwidth]{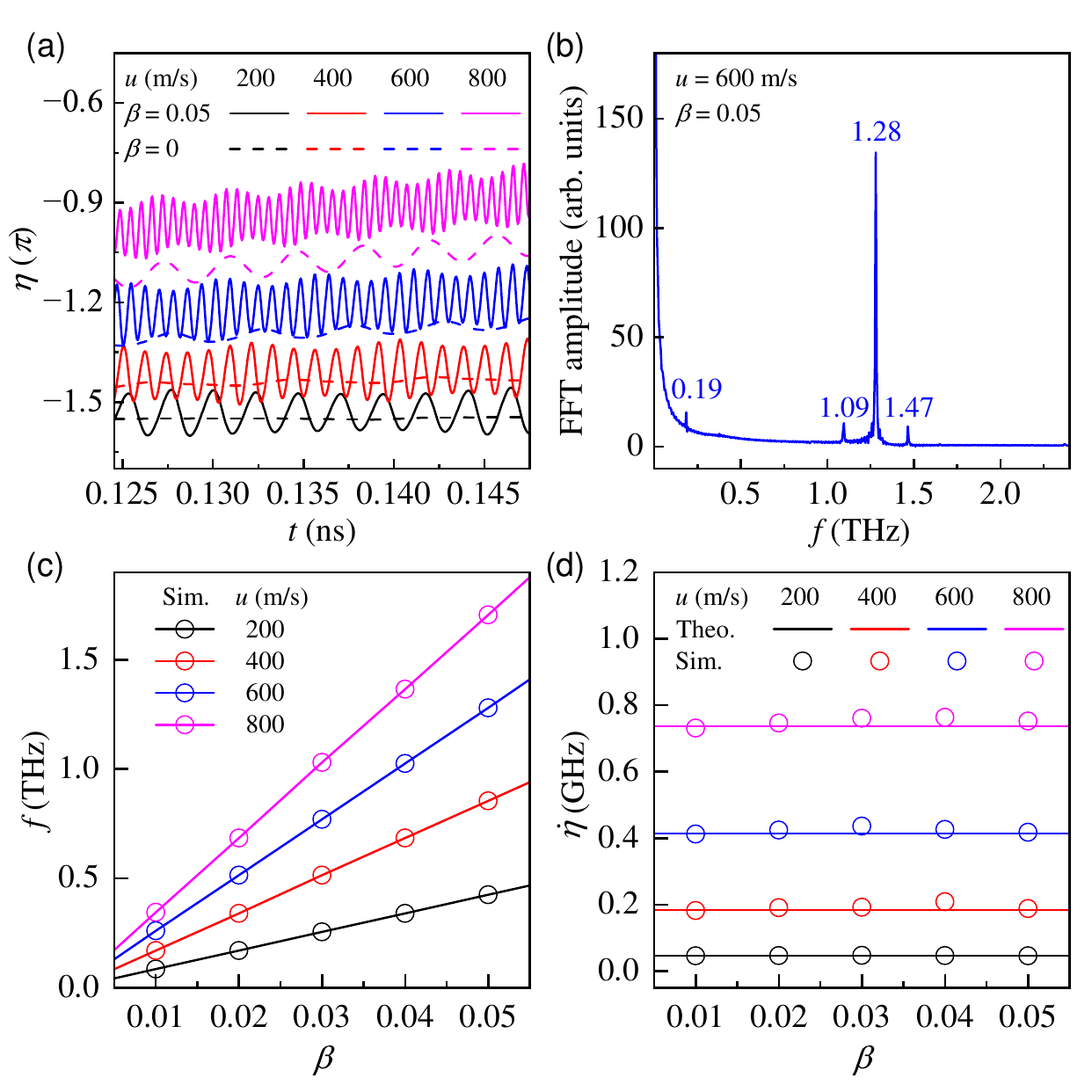}
    \caption{(a) Time-evolution of skyrmion helicity $\eta$ by varying $u$. (b) FFT spectra of the time-evolving skyrmion helicity for $u = 600~\mathrm{m/s}$. Fast (c) and slow (d) frequency component of skyrmion helicity oscillation as a function of $\beta$. In calculations, we consider current flowing along the $\hat{x}$-direction and adopt the following model parameters: $I_{3}=2.4~\mathrm{meV}$, $\alpha=0.05$, and $J_{\mathrm{int}}=4.2~\mathrm{meV}$.}
    \label{fig4}
\end{figure}

In addition to the global sub-GHz helicity rotation, we also observe a THz helicity oscillation [see Fig.~\ref{fig4}(a)]. Surprisingly, the THz mode appears only at finite $\beta$ and its frequency increases linearly with the drift velocity $u$. To interpret this intriguing observation, we extract the time-evolution of the helicity $\eta(t)$ and perform a fast Fourier transformation (FFT). As shown in Fig.~\ref{fig4}(b), the spectrum reveals both fast and slow components. The fast one carries a frequency as high as $1.28~\mathrm{THz}$, which is also captured in the time-evolution of the non-adiabatic force $\mathcal{E}_{\eta,x}$, mirroring the helicity oscillation but with a suppressed amplitude at higher current densities (Sec.~VII~\cite{SM}). It indicates that the fast mode originates from the non-adiabatic torque in Eq.~(\ref{eq9c}). A slow sub-THz helicity oscillation at $0.19~\mathrm{THz}$ is also present. It arises from the $\mathbf{\mathcal{N}}_{\eta}$ term in Eq.~(\ref{eq8b}). The nonlinear mixing between them then generates two sideband modes at $1.09$ and $1.47~\mathrm{THz}$. This is a strong indication of a THz magnon frequency comb \cite{MFC1,MFC2,MFC3}.
To quantitatively explore the role of the non-adiabatic torque, we then simulate the skyrmion motion by varying $\beta$ under a fixed damping parameter $\alpha=0.05$. Figure \ref{fig4}(c) shows that the fast oscillation frequency $f$ grows linearly with $\beta$, confirming its non-adiabatic origin. On the contrary, the global helicity rotation frequency keeps a constant, as shown in Fig.~\ref{fig4}(d), because $\overline{\mathcal{E}_{\eta,x}}$ vanishes.

In the above analysis, we concentrated on the skyrmion motion in nonrelativistic altermagnets. However, relativistic interactions like DMI and DDI are ubiquitous in magnetic materials. They will reshape the skyrmion dynamics, for instance, by setting a periodic pinning potential. In this regard, a finite $\eta$-dependent $V(\eta)$ may emerge, overcoming which requires a large driving current. We envision that the helicity rotation is frozen in the low-current region but exhibits complicated dynamics in the high-current region. Exploring the pinning-depinning is an interesting issue for designing altermagnetic skyrmion qubits in the future. 

\textit{Conclusion}---To summarize, we have studied current-driven skyrmion motion in frustrated altermagnets. We found that the global angular velocity of skyrmion helicity grows quadratically with the current density, and exhibits exotic features of unidirectionality and two-fold anisotropy , in sharp contrast with the well-known ``locked'' helicity picture. Beside the prominent GHz helicty rotation, we also observed a fast THz helicity oscillation purely originating from the non-adiabatic torque. The nonlinear wave mixing eventually generated magnon frequency comb. To obtain analytical understanding, we generalized the Thiele equation to describe the coupled dynamics of skyrmion's center-of-mass and its helicity. Theoretical formulas well explain all numerical findings. Our results uncover a novel DMI-free pathway for nonlinear skyrmion helicity control, distinguishing nonrelativistic altermagnets from conventional magnets, positioning frustrated altermagnets as a transformative platform for skyrmionics, THz technology, magnon frequency comb, and quantum information processing.

\begin{acknowledgments}
This work was funded by the National Key R$\&$D Program of China (No. 2025YFA1411302 and No. 2022YFA1402802), the National Natural Science Foundation of China (NSFC) (No. 12374103 and No. 12434003), and Sichuan Science and Technology Program (No. 2025NSFJQ0045). Y. L. acknowledges the financial support from the China Postdoctoral Science Foundation (Grant No. 2025M773378). Z. J. acknowledges the financial support from NSFC (Grant No. 12404125).
\end{acknowledgments}

\appendix

\nocite{*}

\end{document}